\documentclass[aps,pra,twocolumn,floatfix,longbibliography]{revtex4-1}

\usepackage{graphicx}
\usepackage{amsmath}
\usepackage{amsfonts}
\usepackage{bm}
\usepackage{mathtools}
\usepackage{tikz}
\usepackage{appendix}
\usepackage{ulem}
\usepackage{xcolor}

\begin{document}

\title{Striped Ultradilute Liquid of Dipolar Bosons in Two Dimensions}

\author{Clemens Staudinger$^1$, Diana Hufnagl$^{2,3}$, Ferran Mazzanti$^4$ and Robert E. Zillich$^1$}

\affiliation{$^1$Institute for Theoretical Physics, Johannes Kepler University, Altenbergerstrasse 69, 4040 Linz, Austria
\\ $^2$Johann Radon Institute for Computational and Applied Mathematics, Austrian Academy of Sciences, Altenberger Straße 69, 4040 Linz, Austria
\\ $^3$MathConsult GmbH, Altenbergerstrasse 69, Linz, Austria
\\ $^4$Departament de Física, Campus Nord B4-B5, Universitat Politècnica de Catalunya, E-08034 Barcelona, Spain}

\begin{abstract}
We investigate the phases of a Bose-Einstein condensate of dipolar atoms restricted to move
in a two-dimensional plane.
The dipole moments are all aligned in a direction tilted with respect to the plane normal.
As a result of the attractive and repulsive components 
of the dipole-dipole interaction, the dipolar gas has a self-bound phase,
which is stabilized by quantum fluctuations. Tilting the dipoles tunes the anisotropy 
of the dipole-dipole interaction, which can trigger a spatial 
density modulation. In this work we study these two aspects and investigate the
conditions for the formation of a self-bound and striped phase, which has been realized
in experiments with dipolar droplets.
We use a variational method based on the hypernetted-chain Euler-Lagrange 
optimization of a Jastrow-Feenberg ansatz for the many-body wave function
to study the ground state properties.
This method takes into account quantum fluctuations in a non-perturbative 
way and thus can be used also for strongly correlated systems.
\end{abstract}

\pacs{03.75.Hh, 67.40.Db}

\maketitle

\section{Introduction}

Dipolar quantum gases, and especially dipolar Bose-Einstein condensates (BEC),
are gaining significant attention \cite{chomaz_dipolar_2023}
since self-bound droplets consisting of $^{164}$Dy \cite{luPRL11dysprosium,Kadau2016,Schmitt2016,Ferrier2016, Ferrier2016_2,Ferrier2018}
and $^{166}$Er \cite{chomazPRX16} were realized in experiments.
In contrast to Bose mixtures, the competition between attractive 
and repulsive parts of the interaction does not originate from the 
interaction between the components of the mixture, but rather
from the dipole-dipole interaction itself, which 
in general has repulsive and attractive regions. 
As in Bose mixtures droplets, quantum fluctuations are the driving force behind
the stabilization of dipolar droplets, as confirmed by theory \cite{Wachtler_2016,bailliePRA16,bombinPRL17}. 
More recently, even droplets consisting of dipolar mixtures have been realized
in experiments \cite{Trautmann2018,Durastante2020} and described with 
beyond mean-field methods \cite{Bisset2021,Smith2021}. 
In such droplets the components are not necessarily miscible, but can demix while staying self-bound.

In experiments the dipole moments of all atoms are aligned in parallel 
by an external magnetic field of well-controlled strength
and direction. This provides a means to modify the anisotropy of the dipole-dipole interaction and 
triggers the transition to a density-modulated, self-organized stripe phase, which shows supersolid properties \cite{Leonard2017,Tanzi2019supersolid,Zhang2019,Roccuzzo2019,Hertkorn2021,Tanzi2021science}. 
Such a transition is also visible in the excitation spectrum of a dipolar BEC, where a so-called roton minimum emerges \cite{Santos2003,odellPRL03,chomazNatPhys18,Natale2019,Schmidt2021,Blakie2020}. 
Just like a droplet, a density modulation is a state that is not stable in 
a mean-field approximation~\cite{fischer_stability_2006,Komineas2007}, but rather
stabilized by quantum fluctuations~\cite{Wenzel2018}.

\begin{figure}[h!]
\begin{center}
\includegraphics*[width=0.48\textwidth]{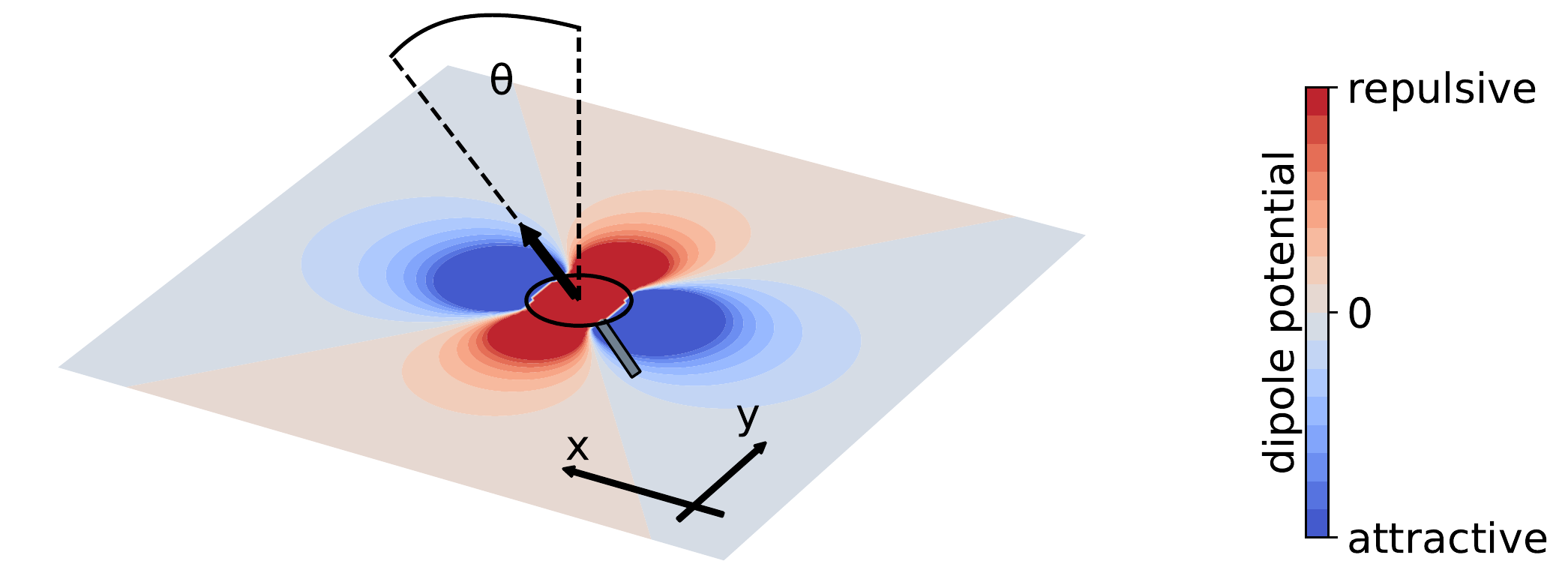}
\end{center}
\caption{
Sketch of the dipolar interaction in 2D,
with tilting angle $\theta$ and repulsion parameter $C_\mathrm{h}$. The direction of the dipole moment 
is shown by the black arrow and the repulsion is indicated by the black circle.
The repulsive and attractive regions of the compound interaction are
shown with red and blue colors, respectively.}
\label{FIG:dipole}
\end{figure}

In previous theoretical studies \cite{maciaPRL12} we observed density modulations 
in the form of stripes in a two-dimensional dipolar
Bose gas with the polarization axis tilted with respect to the
perpendicular direction. In these studies,
where the tilt angle $\theta$ was small enough that
the dipole interaction stayed purely repulsive,
a very high density was required to reach the stripe phase, and no self-binding was involved.
In this work we investigate the formation of 
stripes at much lower densities which can be realized in experiments with magnetic dipole moments.
We achieve this by increasing the tilting angle $\theta$ beyond a critical angle, where the projection of the dipole-dipole interaction
on the 2D plane becomes attractive, see Fig.~\ref{FIG:dipole}, such that self-binding is possible. Density oscillations were observed experimentally \cite{Tanzi2019droplet,Bottcher2019stripes,Chomaz2019,Ilzhofer2021}.
Conventional mean-field theories are not capable of describing such situations,
and more powerful methods like extended mean-field with Lee-Huang-Yang corrections~\cite{Ferrier2016,chomazPRX16}, 
or quantum Monte Carlo (QMC) techniques
\cite{Macia2014,Macia2016,bombinPRL17,Bottcher2019droplet,Kora2019} have to be applied. In this work we employ
the hypernetted-chain Euler-Lagrange (HNC-EL) method \cite{Kro86,KroTrieste,QMBT00Polls}, 
which incorporates correlations beyond
mean-field approaches especially for strongly-correlated and self-bound systems 
\cite{hebenstreitPRA16,Staudinger2018} and requires a much lower computational effort than QMC.

\section{Methodology}

In the following we consider a dipolar Bose gas that is so tightly trapped in the
$z$-direction that we can assume particles are restricted to move in two dimensions,
taken to be the $xy$-plane. The Hamiltonian reads
\begin{align}
    \hat{H} = - \frac{\hbar^2}{2m} \sum\limits_{i=1}^N \Delta_i
    + \sum\limits_{i<j} v(\Vec{r}_i-\Vec{r}_j)~,
\end{align}
with the interaction being the sum of the dipolar term and a respulsive core
\begin{align}
v(\Vec{r}) =\frac{C_\mathrm{{dd}}}{4 \pi}\left[\frac{1}{\left| \Vec{r} \right|^{3}} - \frac{3 \left(x \sin \theta \right)^2}{\left| \Vec{r} \right|^{5}} \right]
+ \frac{E_0 C_\mathrm{h}^{12}}{\left| \Vec{r} \right|^{12}}~. 
\label{eq::dipole-dipole interaction}
\end{align}
In this expression $C_{dd}$ sets the strength of the dipolar interaction and is proportional
to the square of the (electric or magnetic) dipole moment. It is useful to define the characteristic length scale
$r_0=m C_\mathrm{dd}/\left( 4 \pi \hbar^2 \right)$
and the associated energy scale $E_0=\hbar^2/\left( mr_0^2 \right)$, which serve as units for our calculations. 
All dipoles are polarized along a direction in the $xz$-plane that forms an angle $\theta$ with respect to
the $z$-axis, see Fig.~\ref{FIG:dipole}. With this geometry, the dipole-dipole interaction is 
repulsive around the $y$-direction (red regions in Fig.~\ref{FIG:dipole}), but for large enough
$\theta>\theta_\mathrm{c}=\arcsin \left( 1/\sqrt{3} \right) \approx 0.61548$ an attractive region appears in the $x$-direction
(blue regions in Fig.~\ref{FIG:dipole}). In this work we 
explore the highly tilted polarization regime $\theta>\theta_\mathrm{c}$
where the purely dipolar gas is unstable. In order to prevent collapse, we add a
short range repulsive interaction
$C_\mathrm{h}/r^{12}$-potential with the short-range repulsion parameter $C_\mathrm{h}$
as shown in Eq.~(\ref{eq::dipole-dipole interaction}).
As a check of universality of this model we compare the results with those obtained with a $1/r^6$-potential
tuned to the same total scattering length.

We describe the ground state using a variational Jastrow-Feenberg ansatz~\cite{Feenberg} of the form
\begin{align}
    \Psi(\Vec{r}_1, \cdots, \Vec{r}_N) = \exp \left[\frac{1}{2} \sum\limits_{i < j} u(\Vec{r}_{i}-\Vec{r}_{j}) \right] \ ,
    \label{eq:jastrowansatz}
\end{align}
which includes pair correlations $u(\Vec{r})$ and accounts for quantum fluctuations. To obtain the optimal 
ground state we solve the Euler-Lagrange equation
\begin{equation}
{\delta e \over \delta\sqrt{g(\Vec{r})}} = 0
\label{eq:EL}
\end{equation}
where $e$ is the energy per particle
\begin{align}
    e(\rho_0) \equiv \frac{E}{N} = \frac{\rho_0}{2} \int\! \mathrm{d}^2r~ g(\Vec{r}) \left[ v(\Vec{r}) - \frac{\hbar^2}{4m} \nabla^2 u(\Vec{r}) \right]~.
\end{align}
The pair distribution function $g(\Vec{r})$ is given in terms of the wave function 
in Eq.~(\ref{eq:jastrowansatz}) as
\begin{align}
    g(\Vec{r}_1-\Vec{r}_2) = \frac{N(N-1)}{\left<\Psi | \Psi \right> \rho_0^2} \int\! \mathrm{d}^2 r_3\dots \mathrm{d}^2 r_N \left| \Psi(\Vec{r}_1,\dots,\Vec{r}_N) \right|^2,
\end{align}
Closure is provided by the HNC relation between $g(\Vec{r})$ and $u(\Vec{r})$ \cite{Hansen}.
In the following we restrict ourselves to the HNC-EL/0 approximation,
where the so-called elementary diagrams are neglected
in the cluster expansion \cite{Hansen}. We have calculated the leading
contribution of the elementary diagrams to the total energy but found it to be
less than 3\% for densities $\rho_0 r_0^2 \leq 1$ (see appendix \ref{sec::Elementary Diagrams}). 
In the HNC-EL/0 framework, Eq.~(\ref{eq:EL}) can be cast as
\begin{equation}
\Big[-{\hbar^2\over m}\Delta + v(\Vec{r})+w_\mathrm{I}(\Vec{r})\Big]\sqrt{g(\Vec{r})}=0
\label{eq:HNCEL0}
\end{equation}
which has the form of an effective 2-body zero-energy scattering equation
with the bare potential $v$ and an additional induced many-body potential $w_\mathrm{I}$, 
which is defined via its Fourier transform
\begin{equation*}
    w_\mathrm{I}(\Vec{k})=-{\hbar^2k^2\over 4m}\Big(1-{1\over S(\Vec{k})}\Big)^2(2S(\Vec{k})+1)
\end{equation*}
in terms of the static structure factor
\begin{equation}
S(\Vec{k}) = 1 + \mathrm{FT} \left[ g(\Vec{r}) -1 \right]
\label{Sk}
\end{equation}
where $\mathrm{FT}$ denotes the Fourier transformation multiplied with the density $\rho_0$.
We note that Eq.(\ref{eq:HNCEL0}) is not a simple linear differential
equation because the induced potential $w_\mathrm{I}$ depends on $g$ itself. 
The details on how to solve Eq.~(\ref{eq:HNCEL0})
iteratively can be found elsewhere~\cite{KroTrieste,QMBT00Polls}.

From experience with other systems \cite{Clements94,Campbell96,hebenstreitPRA16,raderPRA17,Staudinger2018},
solving the HNC-EL/0 equations is straightforward
for systems with a stable or metastable ground state, but fails to converge if the system is unstable
against infinitesimal perturbations
(e.g. spinodal instability of a system with homogeneous density \cite{Campbell96,Staudinger2018}). Inspection of structural
quantities like $g(\Vec{r})$ and $S(\Vec{k})$ provides clues as to the nature of the instability (e.g.
long-ranged fluctuations in $g(\Vec{r})$ in the case of a spinodal instability). 
More quantitative information on that is provided by a stability analysis of the solution of
the HNC-EL/0 equation~\cite{Castillejo1979}. For this purpose, we evaluate the Hessian, i.e.\
the second functional derivative of the energy $e$ with respect to the pair distribution function,
$K(\Vec{r},\Vec{r}')=\delta^2 e/\delta\sqrt{g(\Vec{r})}\delta\sqrt{g(\Vec{r}')}$.
If this operator is
positive definite, the solution of the HNC-EL/0 equation (\ref{eq:EL}) is stable against infinitesimal
perturbations of $g(\Vec{r})$. This is guaranteed if all eigenvalues $\lambda_i$ in the equation
\begin{equation}
\int\! \mathrm{d}^2r'\,K(\Vec{r},\Vec{r}') f_i(\Vec{r}') = \lambda_i f_i(\Vec{r})
\label{eq:ev1}
\end{equation}
are positive. Conversely, if the lowest eigenvalue $\lambda_0$ is close to zero, the system approaches
an instability. More importantly, the eigenvector $f_0(\Vec{r})$ provides information about the nature
of the instability as shown by our results below.
The explicit form of $K(\Vec{r},\Vec{r}')$ is easily
calculated in the HNC-EL/0 approximation. Following the notation of Ref.~\cite{Castillejo1979}, the
eigenvalue problem can be written as
\begin{equation}
\Big[-{\hbar^2\over m}\Delta + v(\Vec{r})+w_\mathrm{I}(\Vec{r})+\hat W\Big] f_i(\Vec{r}) = \lambda_i f_i(\Vec{r})
\label{eq:ev2}
\end{equation}
where the $\hat W$ operator is defined as
\begin{equation}
\hat W f_i(\Vec{r}) = 
\rho_0 \int\! \mathrm{d}^2r'\,\sqrt{g(\Vec{r})}\,W(\Vec{r}-\Vec{r}')\sqrt{g(\Vec{r}')} f_i(\Vec{r}')
\label{eq:Wdef}
\end{equation}
$W$ is given in in momentum space as
\begin{equation*}
    W(\Vec{k})=-{\hbar^2k^2\over m}\left(1-{1\over S(\Vec{k})^3}\right) \ .
\end{equation*}
Since we only need the lowest eigenvalue to assess the stability, we solve eq.(\ref{eq:ev2}) by
imaginary time propagation, see appendix \ref{sec::Stability Analysis} for details.

\section{Energy and Stability}
\label{sec::Energy and Stability}

In the liquid phase, as opposed to the gas phase, a system is self-bound: the energy per particle
is negative and attains its minimum at an equilibrium density $\rho_\mathrm{eq}$.
Furthermore, below the spinodal density a homogeneous liquid 
becomes unstable against long wavelength density fluctuations, and then
breaks into droplets. In this section we analyze
the ground state energy for various short range repulsion strengths $C_\mathrm{h}$, 
dipole tilt angles $\theta$ and densities $\rho_0$ in order to check whether
the system is in a liquid or in a gas phase.
We also assess the stability against density fluctuations. Instead of reaching
a spinodal instability typical of isotropic liquids, we find a transition to a density
wave in the $y$-direction, i.e.\ a stripe phase.

\begin{figure}[h!]
\begin{center}
\includegraphics*[width=0.48\textwidth]{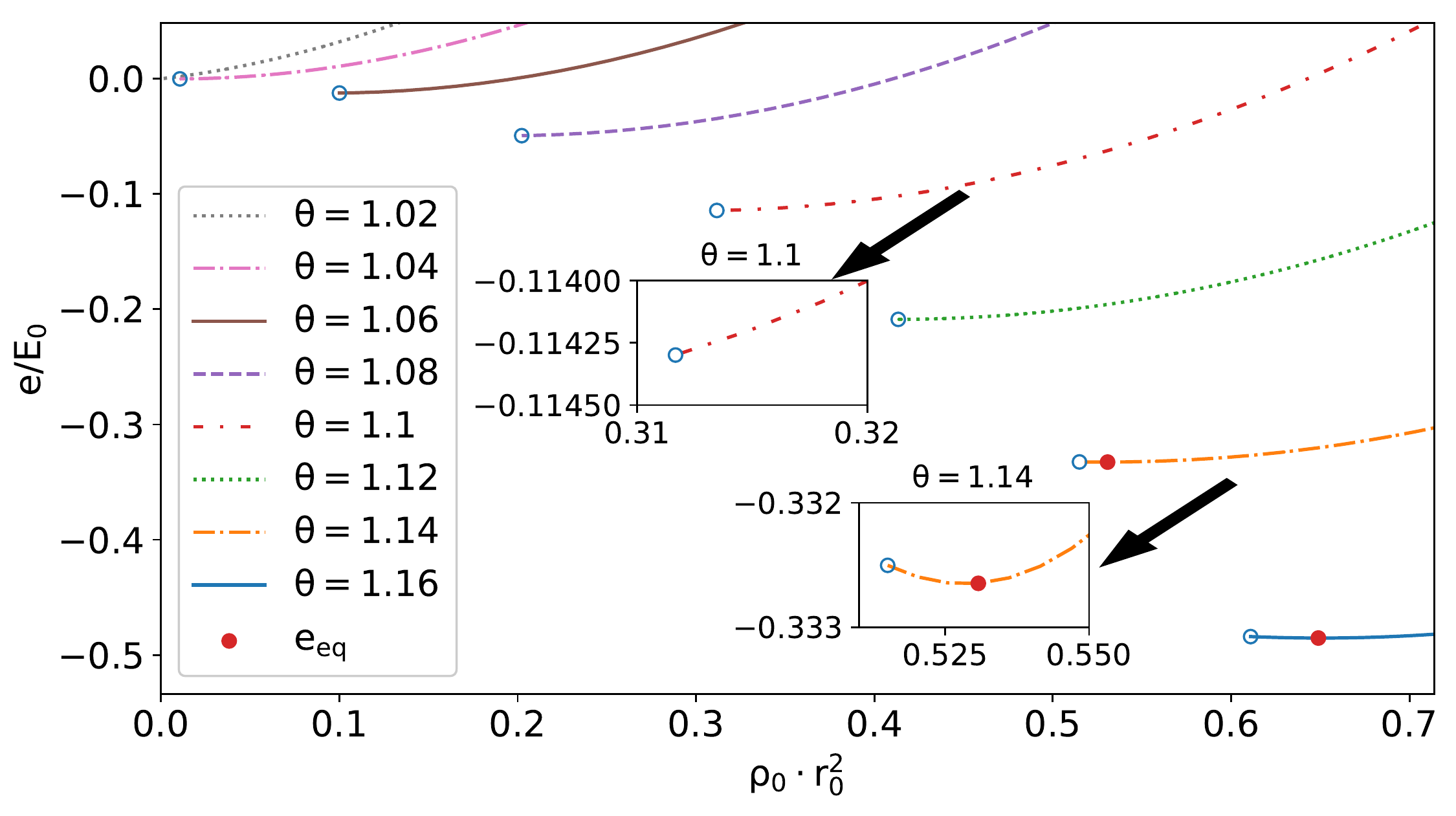}
\end{center}
\caption{Energy per particle $e(\rho_0)$ as a function of the density $\rho_0$ for $C_\mathrm{h}=0.33 r_0$ 
and $\theta=1.02-1.16$. The last converging point at the lowest density in each curve
is marked with a blue open circle and the equilibrium energy $e_\mathrm{eq}$ with a red 
filled circle. The insets show a magnification of the result for $\theta=1.10$ and $1.14$,
which illustrate the behavior of the equation of state near the stripe phase.}
\label{fig::energy cut constant C}
\end{figure}

We first fix $C_\mathrm{h}=0.33 r_0$ and vary the tilt angle between $\theta=1.02$ and $\theta=1.16$. 
The results are depicted in Fig.~\ref{fig::energy cut constant C}, where we show the energy 
per particle $e(\rho_0)$ as a function of the density. 
In each case we start the calculation at a large density and 
solve Eq.~(\ref{eq:EL}) iteratively. The resulting pair distribution function is then used 
as an input to solve the same equation at a lower density, which 
ensures rapid numerical convergence. We repeat this until either reaching zero density or
we don't find a stable solution at non-zero density.
Depending on the tilting angle $\theta$ three different cases can occur, corresponding
to three different phases: a gas, a homogeneous liquid, or a striped liquid.

For $\theta\leq 1.04$ the energy per particle is positive and approches zero as $\rho\to 0$ where
it attains its minimum value $e=0$. The system is then in a gas phase and the corresponding pressure is always positive.
Beyond $\theta=1.04$ the system enters a different phase where the energy per particle $e$ becomes negative as the density is lowered.
The system is self-bound and thus in a liquid phase.  However, as long as $\theta\leq 1.12$, the 
calculation ceases to converge before $e(\rho_0)$ attains a minimum; the density $\rho_\mathrm{c}$
where this happens is indicated with open blue circles in Fig.~\ref{fig::energy cut constant C}.
The inset for $\theta=1.1$ shows this more clearly, where one can see that the liquid phase stops being stable before
reaching a homogeneous equilibrium density $\rho_\mathrm{eq}$
The homogeneous HNC-EL/0 equations are know to cease to converge
at a continuous phase transition~\cite{maciaPRL12,Staudinger2018}.
In the following we show that the dipolar system undergoes a transition to a self-organized stripe phase
where the density exhibits a spatial modulation.

\begin{figure}[h!]
\begin{center}
\includegraphics*[width=0.48\textwidth]{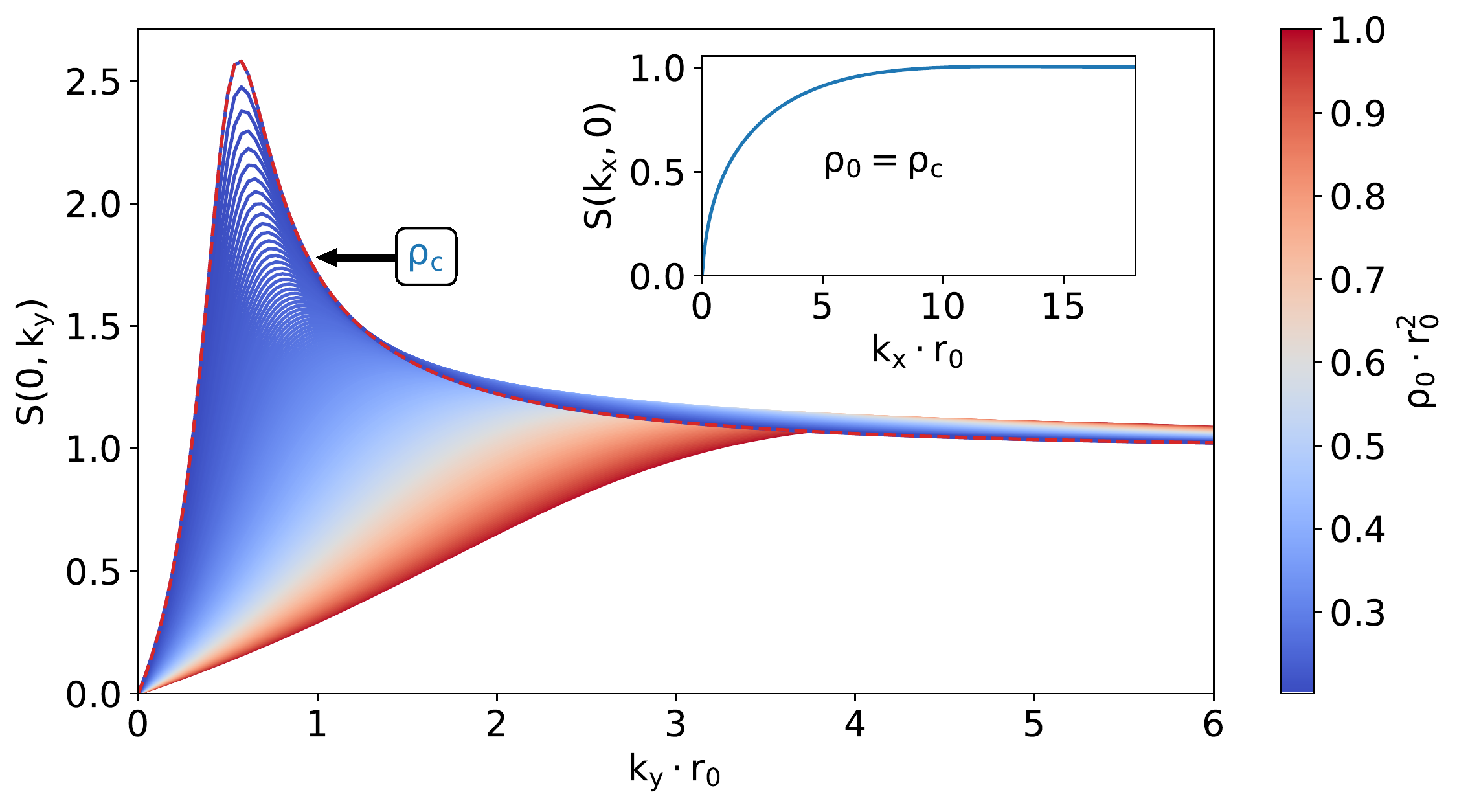}
\end{center}
\caption{Static structure factor $S(0,k_\mathrm{y})$ for $\theta=1.08$ and $C_\mathrm{h}=0.33 r_0$ 
as a function of $k_\mathrm{y}$. 
As the density (color coded) is lowered approaching a critical density 
$\rho_\mathrm{c}=0.2023/r_0^2$, a large peak develops.
$S(k_\mathrm{x},0)$, shown in the inset, is almost independent on $\rho_0$ and does not develop a peak.}
\label{fig::S dipoles stripes}
\end{figure}

\begin{figure}[h!]
\centering
\includegraphics[width=0.48\textwidth]{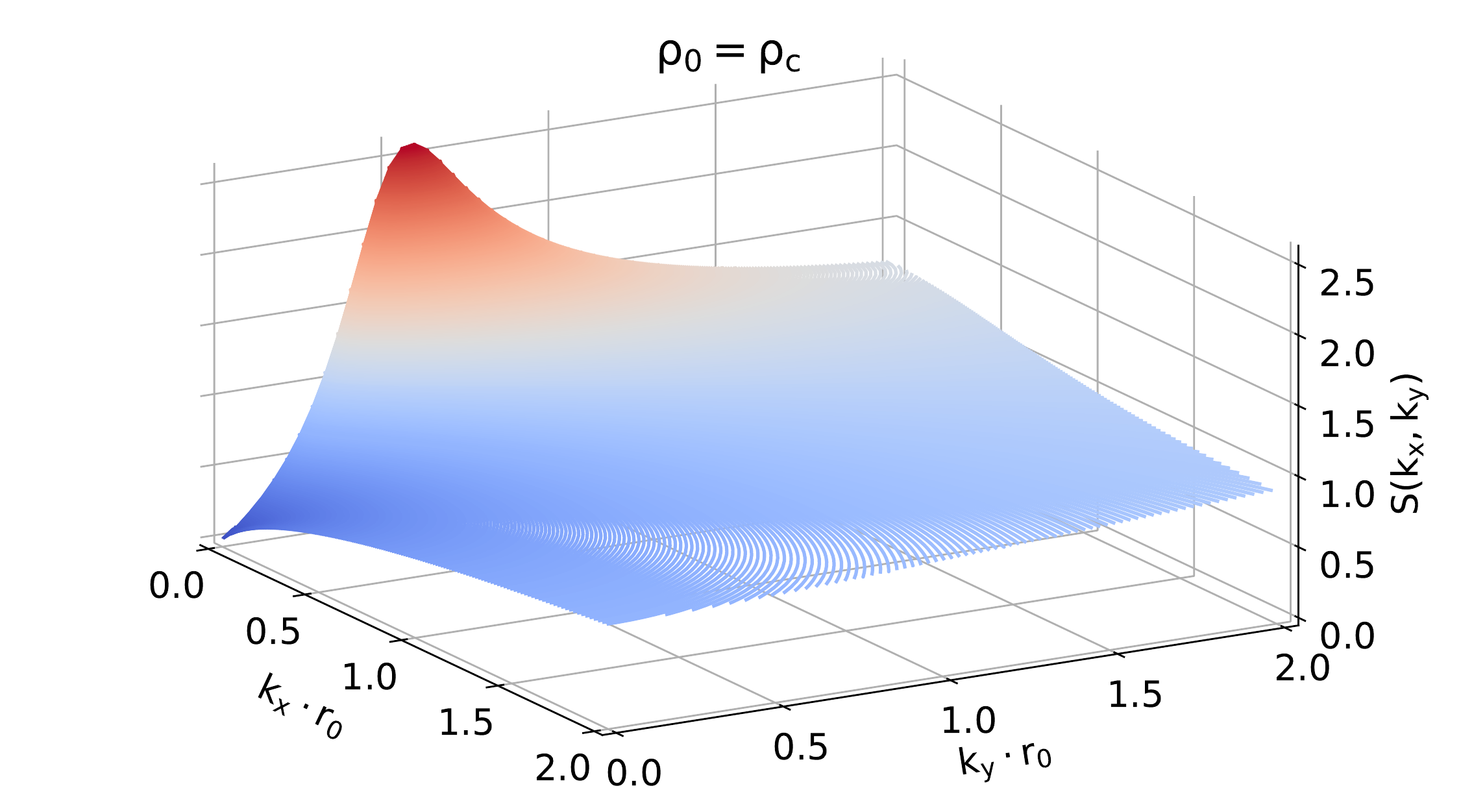}
\caption{Static structure factor $S(k_\mathrm{x},k_\mathrm{y})$ for $\theta=1.08$, $C_\mathrm{h}=0.33 r_0$ 
at the critical density $\rho_\mathrm{c}=0.2023/r_0^2$.}
\label{fig::S3D dipoles stripes}
\end{figure}

\begin{figure}[h!]
\centering
\includegraphics[width=0.48\textwidth]{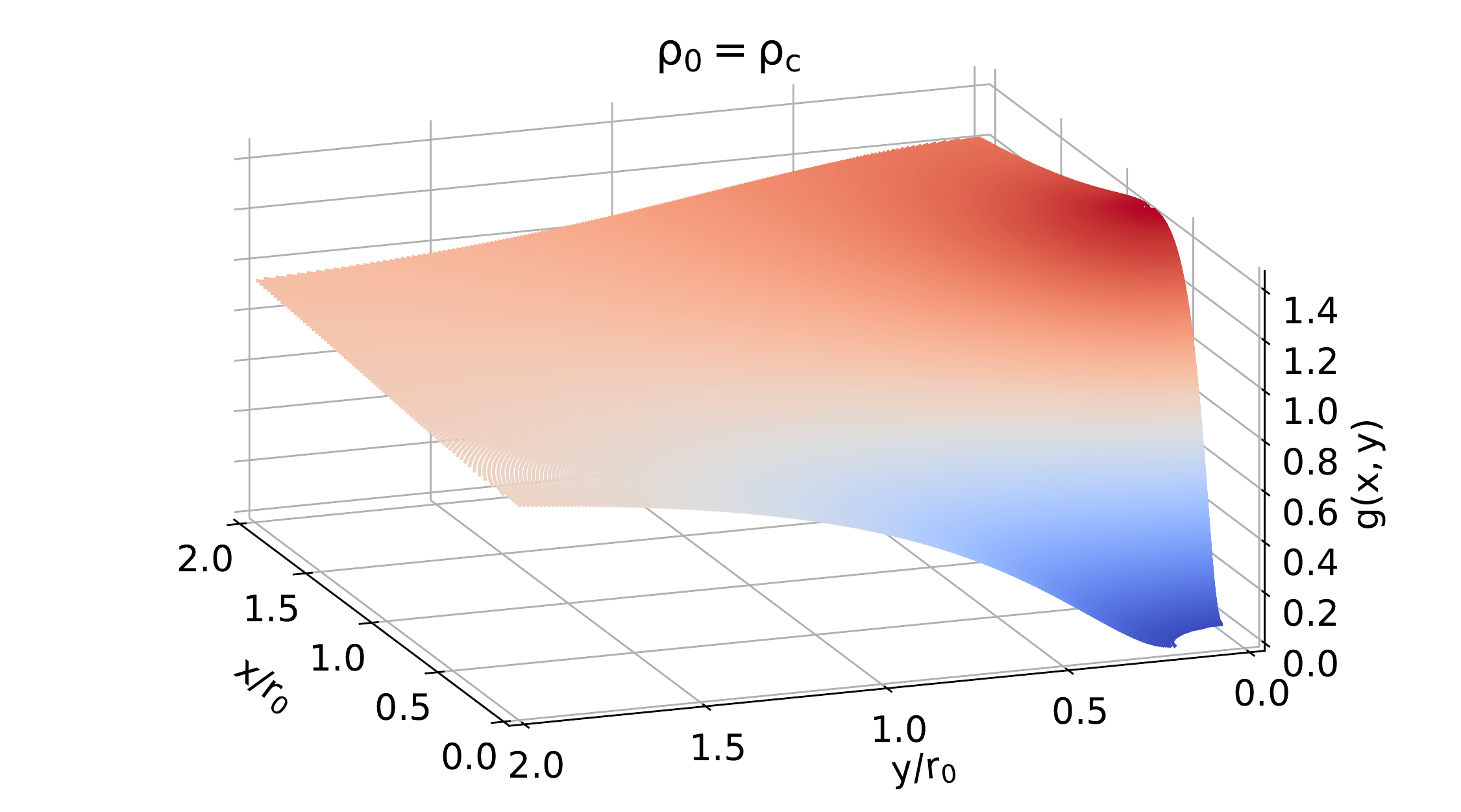}
\caption{Pair distribution function $g(x,y)$ for $\theta=1.08$, $C_\mathrm{h}=0.33 r_0$ at the critical density $\rho_\mathrm{c}=0.2023/r_0^2$.}
\label{fig::g3D dipoles stripes}
\end{figure}

Fig.~\ref{fig::S dipoles stripes} shows the static structure factor $S(k_x=0,k_y)$ along the $y$-direction, 
for $\theta=1.08$ and $C_\mathrm{h}=0.33 r_0$, and densities spanning the range from $0.2023/r_0^2$ to $1/r_0^2$.
A pronounced peak grows as the density approaches a critical value $\rho_\mathrm{c}=0.2023/r_0^2$,
thus signaling increasing spatial ordering between pairs of dipoles in $y$-direction.
This suggests a self-organized long-range order below $\rho_\mathrm{c}$, where homogeneous
HNC-EL/0 does not converge anymore. The same behavior was observed in
Ref.~\cite{maciaPRL12}, except that here the density is three orders of magnitude lower and the dipoles are in a self-bound phase.
From the position $k_\mathrm{p}$ of the peak in $S(k_x=0,k_y)$ at $\rho_\mathrm{c}$
we can predict the wave number of the density oscillation.
The pronounced peak in $S({\bf k})$ is associated with a roton
excitation, according to the Bjil-Feynman approximation for the dispersion relation,
$\hbar\omega(k_x,k_y)={\hbar^2 k^2\over 2m\,S(k_x,k_y)}$,
which is expected to work well at low densities. In this way,
the reported structure factor points to the emergence of a roton instability along 
the direction of $k_y$, compatible with a transition to a stripe phase.
In contrast to the $y$-direction, the structure factor in $x$-direction 
has no peak (see inset in Fig.~\ref{fig::S dipoles stripes}), 
thus showing no signs of ordering in $x$-direction.
Figure~\ref{fig::S3D dipoles stripes} shows the full $S(k_\mathrm{x},k_\mathrm{y})$ at the critical density $\rho_c$.

Fig.~\ref{fig::g3D dipoles stripes} shows the corresponding pair distribution function $g(x,y)$
obtained as the Fourier transform of $S(k_x,k_y)$. As expected $g(x,y)$ shows a small peak 
along the $x$ direction where the attractive well of the dipole-dipole interaction (Fig.~\ref{FIG:dipole}) is deepest.
Other than that, the pair distribution function is smooth.
In the small range depicted, $g(x,y)$ has little structure along the $y$-direction,
and the fluctuations in $g(0,y)$ that lead to the peak in $S(0,k_y)$ cannot be seen.

\begin{figure}[h!]
\begin{center}
\includegraphics*[width=0.48\textwidth]{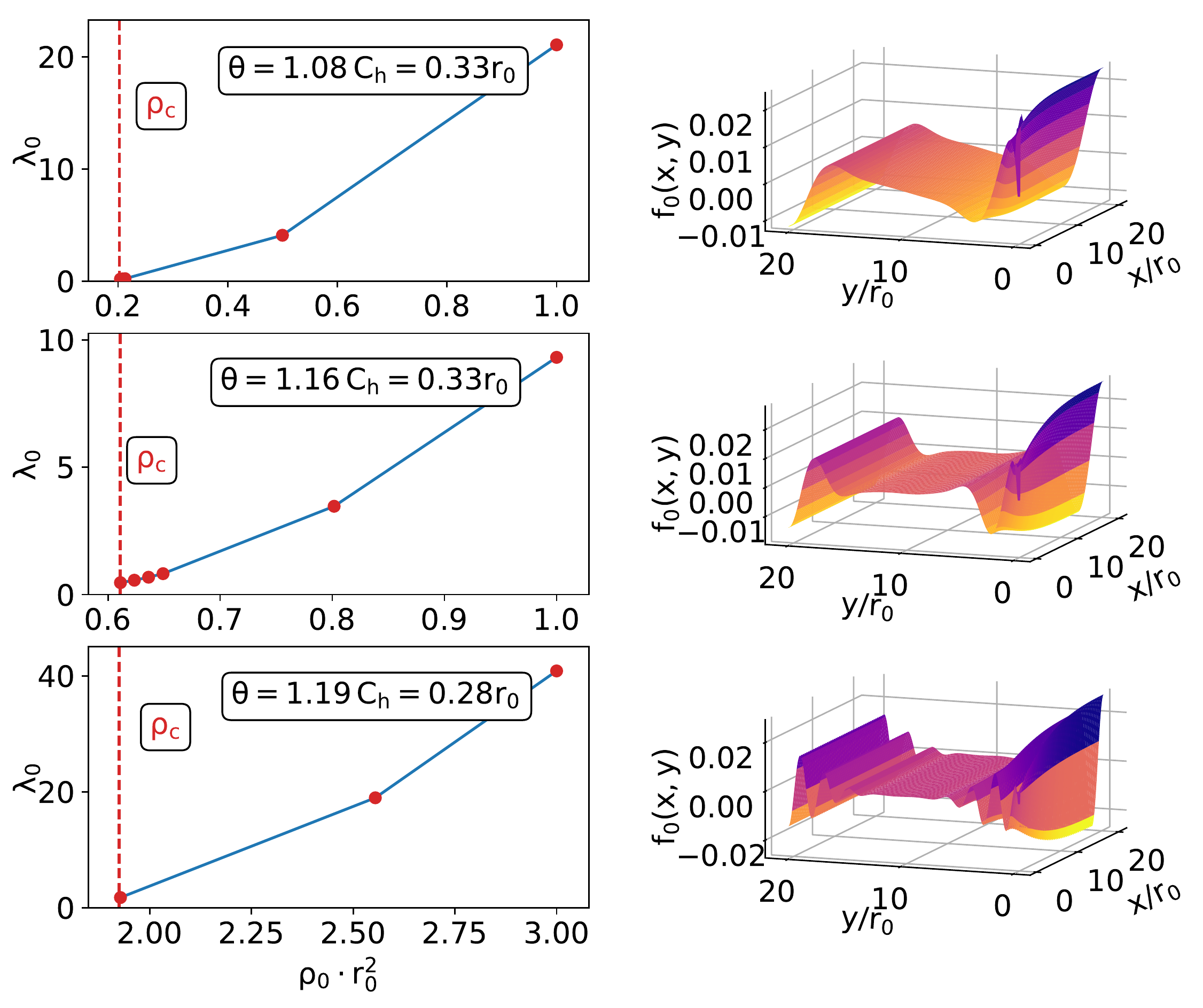}
\end{center}
\caption{Stability analysis for $C_\mathrm{h}=0.33 r_0$, $\theta=1.08/1.16$ and $C_\mathrm{h}=0.28 r_0$, $\theta=1.19$. 
The left column shows the lowest eigenvalue $\lambda_0$ 
of the Hessian as a function of the density $\rho_0$, while the right colum shows
the eigenvector $f_0(\Vec{r})$ at $\rho_\mathrm{c}$.}
\label{fig:stability analysis}
\end{figure}

The emergence of a peak in the static structure factor along the
$y$-direction is thus a strong indicator for the transition to a stripe phase. 
A more rigorous stability analysis explained above leads to the same conclusion, as illustrated in 
Fig.~\ref{fig:stability analysis}. The top, middle and bottom panels correspond to
($C_\mathrm{h}=0.33 r_0$, $\theta=1.08$),
($C_\mathrm{h}=0.33 r_0$, $\theta=1.16$) and
($C_\mathrm{h}=0.28 r_0$, $\theta=1.19$), respectively. The left panels depict the density dependence of
the lowest eigenvalue $\lambda_0$ of Eq.(\ref{eq:ev2}).  
As can be seen, $\lambda_0$ approaches zero as the density approaches $\rho_\mathrm{c}$, confirming 
that the homogeneous phase becomes unstable in that limit. As a further confirmation
we show the lowest eigenvectors $f_0(\Vec{r})$ at $\rho_\mathrm{c}$ in the right column of the same figure.
The shape of $f_0$ informs about the least stable fluctuation in
$g(x,y)$, which drives the transition to a stripe phase,
showing oscillations along the $y$-direction. These oscillations are most pronounced
at high densities, as seen in the bottom panel of the figure. 
The wave number of this oscillation is the same as the wave number $k_\mathrm{p}$
of the peak in $S(0,k_y)$ at $\rho_\mathrm{c}$ (see Fig.~\ref{fig::S dipoles stripes}).
This behavior clearly signals the transition to a phase with long range order in the $y$-direction.  
Note that in all cases, the oscillation is strongly damped, which
was not found for the stripe phase transition at high densities
in Ref~.\cite{maciaPRL12}, where the system is not self-bound.

When the tilting angle is increased beyond $\theta= 1.12$ while still keeping $C_\mathrm{h}=0.33 r_0$, the energy
decreases further due to the stronger attraction, but there is also a qualitative change of the shape of the equation
of state shown in Fig.~\ref{fig::energy cut constant C}: $e(\rho_0)$ is not monotonous anymore, but reaches a minimum at
a homogeneous equilibrium density $\rho_\mathrm{eq}$ before reaching the critical density $\rho_c$ of the transition to the stripe phase.
This new minimum corresponds to a self-bound homogeneous liquid state. If the system were
finite, the dipoles would form a two-dimensional droplet adjusting its radius so
that the pressure inside the droplet is zero at $\rho_\mathrm{eq}$. 
When the density is lowered further the energy per particle starts
increasing again, up to the point where the transition to the stripe phase takes place at $\rho_\mathrm{c}$,
as evidenced by the stability analysis shown in Fig.~\ref{fig:stability analysis}).
The lower inset in Fig.~\ref{fig::energy cut constant C} shows the result for $\theta=1.14$, 
where the energy minimum at $\rho_\mathrm{eq}$ is clearly visible.

\section{Phase Diagram}

\begin{figure}[h!]
\begin{center}
\includegraphics*[width=0.48\textwidth]{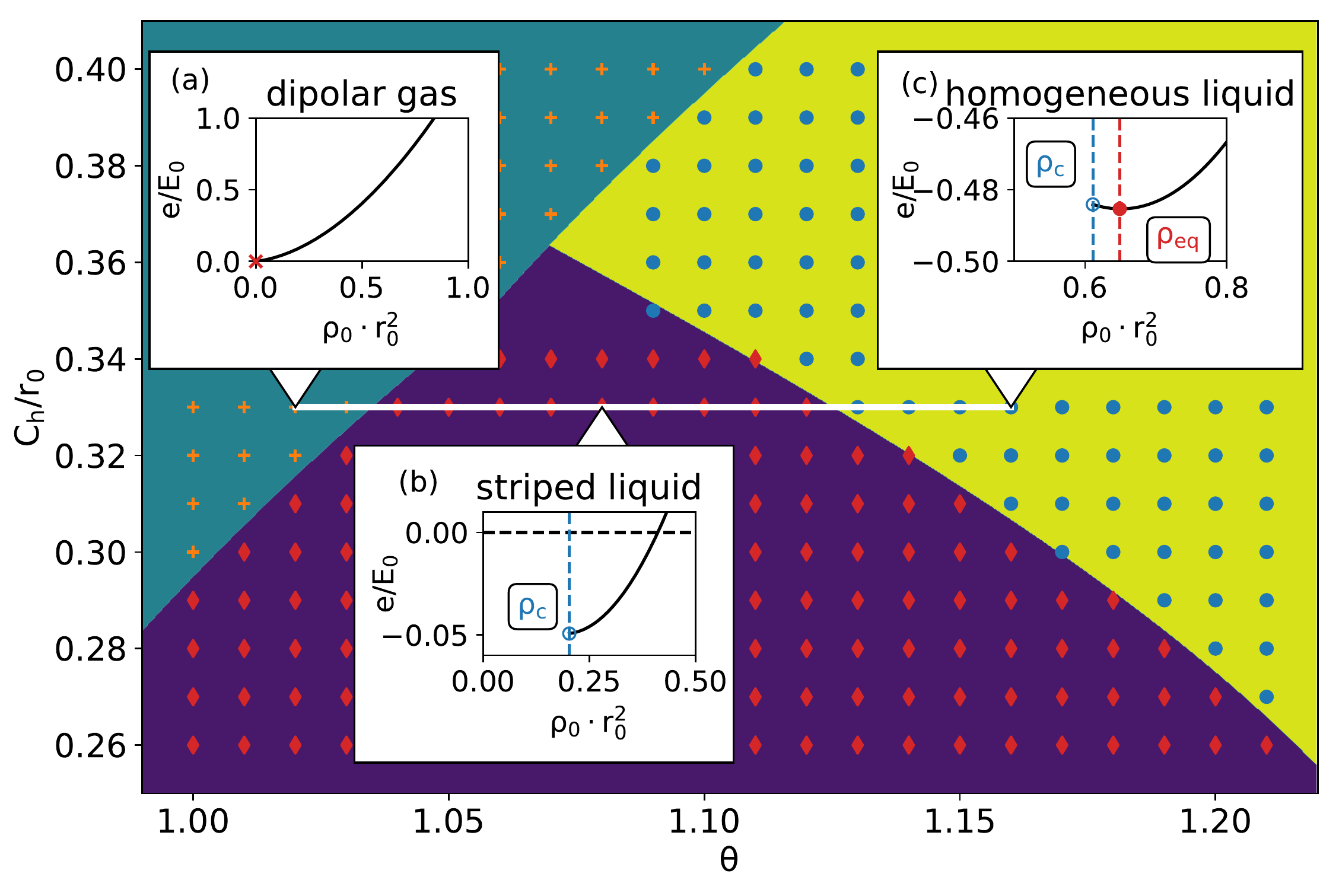}
\includegraphics*[width=0.48\textwidth]{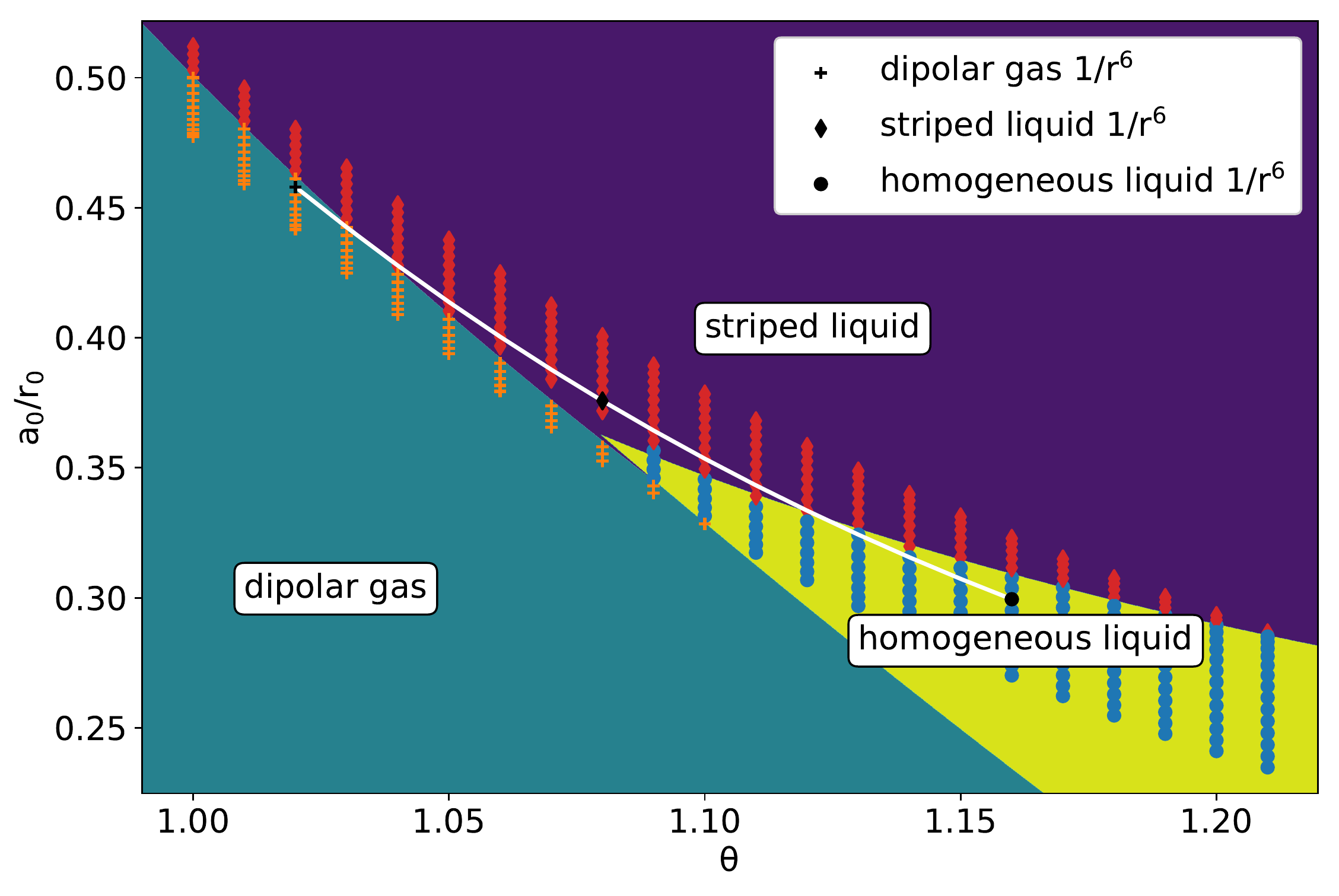}
\end{center}
\caption{
Phase diagram as a function of the
tilting angle $\theta$ and short-range repulsion $C_\mathrm{h}$ (upper panel) and 
as a function of the scattering length $a_0$ (lower panel). The insets (a), (b), and (c) 
show a representative $e(\rho_0)$ for each phase for $\theta=1.02$, $1.08$, and $1.16$, respectively,
all for the same value $C_\mathrm{h}=0.33 r_0$ (white line).
In the lower panel the three black symbols show the result of a calculation with a $1/r^6$-repulsion, which gives
the same result as the $1/r^{12}$-repulsion used for all other calculations.}
\label{fig:Phase Diagram}
\end{figure}

\begin{figure}[h!]
\begin{center}
\includegraphics*[width=0.48\textwidth]{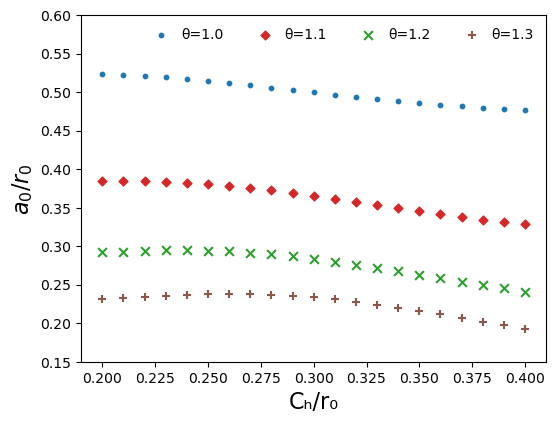}
\end{center}
\caption{
Scattering length $a_0$ of the interaction (\ref{eq::dipole-dipole interaction}) as function of the
repulsion parameter $C_h$ for 4 different tilting angles $\theta$.}
\label{fig:Scatt_Length}
\end{figure}

In order to obtain the full phase diagram, we calculate $e(\rho_0)$ for a range of $\theta$ and $C_\mathrm{h}$
values following the protocol described above, i.e. lowering $\rho_0$ from a sufficiently large value down to zero or
until a homogeneous phase ceases to be stable at a finite density value $\rho_\mathrm{c}$.
We analyze the results in the same way as in the previous section 
to classify the phase into gas, striped liquid, and homogeneous liquid.
Fig.~\ref{fig:Phase Diagram} presents this classification, indicating the phases by different colors,
as function of $\theta$ and $C_\mathrm{h}$: evaporating gas,
self-bound striped liquid,
or self-bound liquid with no stripes (with equilibrium density $\rho_\mathrm{eq}$).
The circles show the grid of $\theta$ and $C_\mathrm{h}$ values that we used for the classification.
The lower panel of Fig.~\ref{fig:Phase Diagram} depicts the same phase diagram as the upper panel, but
in terms of the scattering length $a_0$ of the interaction (\ref{eq::dipole-dipole interaction}) instead of $C_\mathrm{h}$ 
to make it independent of the model for the repulsion~\cite{maciaPRA11}.
The scattering length $a_0$ depends on both $C_\mathrm{h}$ and $\theta$, and increasing $C_\mathrm{h}$ decreases $a_0$,
see Fig.~\ref{fig:Scatt_Length}.
We also include three points, marked in black, corresponding to an 
interaction $1/r^6$ but tuned to give the same scattering length of the compound system.
These calculations give the same phase as those with the $1/r^{12}$-repulsion,
which shows our results are universal, independent of the 
model for the short-range interaction.
Based on the grid of $\theta$ and $C_\mathrm{h}$ values,
the boundaries between the phases were calculated using a C-support vector machine \cite{scikit-learn}, and
an analytic expression for those phase boundaries is given in appendix \ref{sec::phase boundaries}.
In the upper panel of Fig.~\ref{fig:Phase Diagram} a white line is drawn for the fixed repulsion parameter
$C_\mathrm{h}=0.33 r_0$ that we used in most calculations presented
in the previous section. The three insets show three examples for the equation
of state $e(\rho_0)$ along this line, each representing one of the phases. The points along the white line are
the results based on the energies $e(\rho_0)$ shown in Fig.~\ref{fig::energy cut constant C}.
Note, that a dipolar system in 2D can also have density oscillations at very high densities $\rho_0 \sim 10^2/r_0^2$ 
according to Ref.~\cite{maciaPRL12}. However, here we focus on 
the formation of a stripe phase at orders of magnitude lower densities.
The critical densities below which we predict a stripe phase,
are discussed in the following section.

Fig.~\ref{fig:Phase Diagram} shows that,
for tilting angle $\theta>1.07$, upon increasing the repulsion strength $C_h$
the system undergoes first a transition from a striped liquid to a homogeneous liquid, and then 
a transition to a homogeneous gas. The reason is that the
increasing isotropic short-range repulsion becomes dominant compared to the anisotropic dipole-dipole interaction, 
which is responsible for the formation of stripes. 
More interestingly, an increase of the tilting angle $\theta$, and thus of the anisotropy, 
for a fixed short-range interaction, also drives the system to the homogeneous liquid phase,
as we have discussed in Sec.~\ref{sec::Energy and Stability}.
This might seem counter-intuitive at first. However, it
is actually the repulsive part of the dipole-dipole interaction in
$y$-direction causing the formation of stripes and this part of the 
interaction does not change with $\theta$.
At the same time the attraction in $x$-direction increases with increasing $\theta$ and leads to a more strongly bound
system with a higher density, and the stripes merge as their wave length decreases. For example if $C_\mathrm{h}=0.33r_0$
is fixed and the tilting angle is increased from $\theta=1.08$ to $\theta=1.16$ the wave length decreases from
$\lambda=11.70r_0$ to $\lambda=4.82r_0$.

The phase diagram in the lower panel of Fig.~\ref{fig:Phase Diagram}, where
we characterize the phase diagram using the scattering length $a_0$ of the full interaction
the phase diagram, is more intuitive: for a given $a_0$, the system undergoes a transition from an homogeneous
liquid to to a striped liquid phase with increasing $\theta$.

\section{Critical Density}

\begin{figure}[h!]
\begin{center}
\includegraphics*[width=0.48\textwidth]{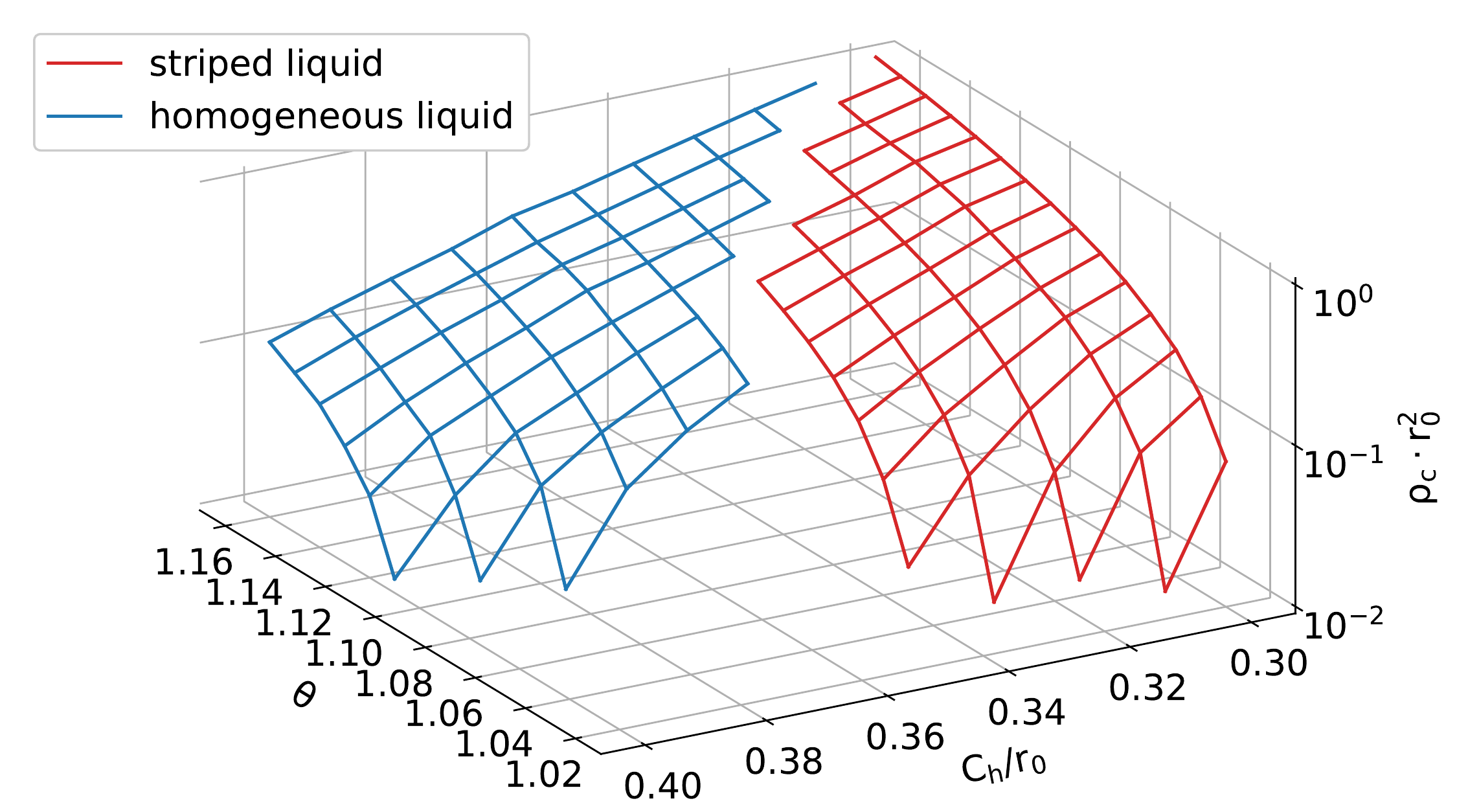}
\end{center}
\caption{Critical density $\rho_\mathrm{c}$ for the transition to the stripe phase as a function of $\theta$ and $C_\mathrm{h}$.
The red and blue surfaces denote the regimes of striped liquid and homogeneous liquid, respectively, see Fig.~\ref{fig:Phase Diagram}.
At the border to the gas phase, $\rho_\mathrm{c}$ falls to zero.}
\label{fig::critical density}
\end{figure}

Figure~\ref{fig::critical density} shows the critical density $\rho_\mathrm{c}$ for the transition to a 
stripe phase as a function of tilting angle $\theta$ and repulsion strength $C_\mathrm{h}$, showing
that $\rho_\mathrm{c}$ increases with increasing $\theta$ and decreasing $C_\mathrm{h}$. 
The blue and red surfaces represent the regions of the homogeneous liquid and 
the striped liquid, respectively, see the phase diagram in Fig.~\ref{fig:Phase Diagram}. Both
surfaces approach $\rho_\mathrm{c}=0$ as the system approaches the gas phase, seen as a sharp
drop on the logarithmic scale in Fig.~\ref{fig::critical density}.
This shows that with a proper choice of
$\theta$ and $C_\mathrm{h}$ (or $a_0$), stripes can be observed in arbitrarily dilute systems.
This is markedly different
from previous theoretical studies of tilted dipoles in two dimensions \cite{maciaPRL12}, where the dipole orientation was not
tilted above $\theta=\arcsin(1/\sqrt{3})$ such that the dipole-dipole interaction stays purely repulsive.
In that case a stripe phase was predicted only above a critical density $\rho_\mathrm{c}=\mathrm{O}(10^2)$. 
We also note that in the range of $\theta$ and $C_\mathrm{h}$ explored in this work,
we have $\rho_\mathrm{c}r_0^2<1$ . Despite the variational nature of the calculations presented, 
the contribution of elementary diagrams to the energy per particle
is less than 3\%\ (see appendix~\ref{sec::Elementary Diagrams} for details), 
while three- or higher-order correlations are negligible in the 
low density regime considered.

\section{Conclusion}

In this work we have explored the phase diagram of a dipolar gas in two dimensions as a function of the
dipole tilting angle and their short-range repulsion. 
We have found that this system exhibits three different
phases: a gas phase, a self-bound stripe phase, and a homogeneous self-bound stripeless phase.
We detected the transition to a stripe phase by monitoring a peak that emerges in the structure factor 
close to a critical density $\rho_c$ that marks the phase transition. 
The critical density depends on the short range repulsion and the tilting angle, and
can be low enough so that it can be realized with current experimental setups.
To further confirm the emergence of the low density stripe phase we have also
conducted a stability analysis based on the Hessian of the energy.
Although our calculations, based on the variational HNC-EL/0 method, always assumed homogeneity, this 
provides strong evidence that a self-bound and striped droplet can form in a
two-dimensional anisotropic dipolar Bose gas.

While our results are compatible with a continuous transition from a normal dipolar liquid to
a self-bound stripe phase, results for the latter phase are necessary in order to 
understand the properties of this phase and the nature of the phase transition,
be it first or second order.
For this purpose one can use a variational model that allows for density modulations, such
as the inhomogeneous generalization of the HNC-EL method~\cite{kroValencia98},
which has been used for example to describe dipolar systems in quasi-2D
geometries~\cite{hufnaglPRL11,hufnaglPRA13}. Alternatively 
one can use exact diffusion Monte Carlo methods, possibly with optimal HNC-EL solutions as
guiding wave functions.  We are pursuring the latter option
to study the self-bound nature of two-dimensional striped dipolar systems.

\begin{acknowledgments}
We acknowledge discussions with Jürgen T. Drachta and Tilman Pfau.
F. Mazzanti acknowledges financial support by grant PID2020-113565GB-C21 funded by MCIN/AEI/10.13039/501100011033, 
and from Secretaria d'Universitats i Recerca del Departament d'Empresa i Coneixement de la
Generalitat de Catalunya, co-funded by the European Union Regional Development
Fund within the ERDF Operational Program of Catalunya (project QuantumCat, ref.
001-P-001644).
\end{acknowledgments}

\appendix

\section{Stability Analysis}
\label{sec::Stability Analysis}

Solving for the lowest eigenvalue $\lambda_0$ and associated eigenvector $y_0(x,y)$ of Eq.~(\ref{eq:ev2})
can be done in several ways. Here we use imaginary time propagation
\begin{equation}
    f_0 = \lim\limits_{t\to\infty}\exp\Big[-t\Big(-{\hbar^2\over m}\Delta + v(\Vec{r})+w_\mathrm{I}(\Vec{r})+\hat W\Big)\Big] \tilde f
\end{equation}
where $\tilde f$ is an initial guess with the only requirement that it has a non-zero projection
onto $f_0$. As usual, the time propagation is split into small time
steps $\tau$ such that the exponentiation can be approximated in a suitable way. Here
we use the Trotter approximation~\cite{trotter59}. Taking the exponent of the potential
functions $v(\Vec{r})$ and $w_\mathrm{I}(\Vec{r})$ is trivial in $r$-space, 
while the exponential of the kinetic part is carried out in momentum space.
However, the integral operator $\hat W$ cannot be easily exponentiated.  We could take the further approximation,
$\exp[-\tau W]\approx 1-\tau\hat W$, but this would require an extremely small
time step $\tau$. Therefore, we split the integration kernel in (\ref{eq:Wdef}) as
$\hat W=\hat W_1+\hat W_2+\hat W_3$ with
\begin{align*}
    \hat W_1\, f_i(\Vec{r}) &=
    \rho_0 \int\! \mathrm{d}^2r'\,(\sqrt{g(\Vec{r})}-1)W(\Vec{r}-\Vec{r}')\sqrt{g(\Vec{r}')} f_i(\Vec{r}')\\
    \hat W_2\, f_i(\Vec{r}) &=
    \rho_0 \int\! \mathrm{d}^2r'\,W(\Vec{r}-\Vec{r}')(\sqrt{g(\Vec{r}')}-1) f_i(\Vec{r}')\\
    \hat W_3\, f_i(\Vec{r}) &=
    \rho_0 \int\! \mathrm{d}^2r'\,W(\Vec{r}-\Vec{r}') f_i(\Vec{r}')
\end{align*}
The first two terms turn out to be small because $\sqrt{g(\Vec{r})}-1$ becomes small for large $r$.
The third term is still large but has the form of a convolution integral, hence it can be easily
exponentiated in momentum space. We obtain the following approximate imaginary time propagation operator
for small time steps $\tau$
\begin{align*}
  G =&
  e^{-{\tau\over 2}[{\hbar^2k^2\over m}+W(\Vec{k})]}
  \ {\cal F}\
  e^{-{\tau\over 2}[V(\Vec{r})+w_I(\Vec{r})]}
  \big[1-\tau (\hat W_1+\hat W_2)\big]\\
& e^{-{\tau\over 2}[V(\Vec{r})+w_I(\Vec{r})]}
  \ {\cal F}^{-1}\
  e^{-{\tau\over 2}[{\hbar^2k^2\over m}+W(\Vec{k})]}
\end{align*}
where we used a symmetric form of the Trotter approximation. ${\cal F}$ denotes the fast Fourier transformation.
For this propagator we can use a time step two orders of magnitude larger than for a propagator
without the above splitting of $\hat W$ into small and large contributions.

\section{Phase Boundaries}
\label{sec::phase boundaries}

\begin{table}[h!]
    \centering
    \begin{tabular}{c|c c c c c c}
        $y_j^i \alpha_{ij}$  \\ \hline
        $j=1$ &0. &0.776  &0. &0. &0. &2.607 \\
        &6. &0. &6. &6. &2.872  &0. \\ \hline
        $j=2$ &0. &0. &-3.384 \\
        &0.354  &0.770  &0. \\ \hline
        $j=3$ &0. &-2.775 &-6. &-2.285 &-6. &-3.812 \\
        &-1.124 &0. &0. &0. &0.  &0.
    \end{tabular}
    
\flushleft
    \begin{tabular}{c|c c}
         & $\theta$ & $C_\mathrm{h}/r_0$  \\ \hline
        SV class 1 & 1.14 & 0.32 \\
        & 1.05 & 0.34 \\
        &1.11  &0.34 \\
        &1.18 &0.29 \\
        &1.19 &0.28 \\
        &1. &0.29 \\ \hline
        SV class 2 &1.08 &0.38 \\
        &1.09 &0.39 \\
        &1.01 &0.31 \\ \hline
        SV class 3 & 1.09 &0.38 \\
        & 1.13 &0.33 \\
        &1.09 &0.35 \\
        &1.16 &0.31 \\
        &1.17 &0.3 \\
        &1.21 &0.27
        \end{tabular}
    \caption{Dual coefficients and support vectors (SV) for the upper panel of Fig.~\ref{fig:Phase Diagram}.}
    \label{tab::parameter upper panel}
\end{table}
We use a C-support vector machine~\cite{scikit-learn} to 
trace the boundaries of the different phases shown in Fig.~\ref{fig:Phase Diagram}. 
We have trained a classifier for each pair of the three classes ($j=1$ for striped liquid, $j=2$ for dipolar gas,
and $j=3$ for homogeneous liquid) The classifiers have the form
\begin{equation}
    f_j(\Vec{x})= \sum\limits_i y_j^i \alpha_{ij} k(\Vec{x},\Vec{x}^i)+b_j~.
\end{equation}
where $k$ is the kernel, which has been chosen to be simply
\begin{equation}
    k(\Vec{x},\Vec{x}')=\left( \gamma \langle \Vec{x},\Vec{x}' \rangle \right)^d~,
\end{equation}
where $\langle \dots \rangle$ denotes the dot product and $\Vec{x}=(\theta, C_\mathrm{h})$.
For the upper panel in Fig.~\ref{fig:Phase Diagram}, $d=6$,  $\gamma=3.2531$ and $b_j= \lbrace -27.5267, 56.9545, 23.9379 \rbrace$.
The dual coefficients $y_j^i \alpha_{ij}$ and the support vectors (SV) $\Vec{x}^i$ are found in tab.\ I.
Each support vector is used in two classifiers so there are two dual coefficients for each $\Vec{x}^i$
(columns in the upper part of tab.\ I). The phase boundaries between class $j$ and $k$ are then obtained 
imposing the condition
\begin{equation}
    f_j(\Vec{x})-f_k(\Vec{x})=0~.
\end{equation}
The classifier was trained with ten-fold cross-validation and a regularization parameter of $C=6$, which avoids over-fitting
of the training data.

\section{Elementary Diagrams}
\label{sec::Elementary Diagrams}

\begin{table}[h!]
    \centering
    \begin{tabular}{c|c|c|c}
        $\theta$ & $C_\mathrm{h}/r_0$ & $\rho_\mathrm{c/eq} \cdot r_0^2$ & $\Delta e_\mathrm{ele} [\%]$  \\ \hline
        1.08 &0.33 &0.2023 &0.01 \\
        1.16 &0.33 &0.6489 &0.93 \\
        1.14 &0.30 &1.00 &2.94 \\
        1.19 &0.28 &1.9245 &9.31 \\
        1.21 &0.27 &2.4742 & 13.95 \\
        1.21 &0.26 &2.9948 &20.68 \\
    \end{tabular}
    \caption{Contribution of the elementary diagrams to the total energy $\Delta e_\mathrm{ele}$ in percent. The third column either gives the critical density $\rho_\mathrm{c}$ (rows 1, 3, 4 and 6) or the equilibrium density $\rho_\mathrm{eq}$ (row 2 and 5).}
    \label{tab::elementary diagrams contribution}
\end{table}

We investigated the influence of elementary diagrams by calculating the energy contribution $e_\mathrm{ele}$
of the lowest order elementary diagram, the four point diagram \cite{Hansen}. We calculate $e_\mathrm{ele}$ using the pair
distribution function $g(x,y)$ of the HNC-EL/0 results, i.e.\ from the energy optimization without elementary
diagrams. This is justified if the contribution of elementary diagrams is small, otherwise elementary diagrams have to
be included self-consistently in the optimization. We report the relative change in energy compared to the HNC-EL/0 result
\begin{equation}
    \Delta e_\mathrm{ele} = \frac{e_\mathrm{ele}}{e+e_\mathrm{ele}}~.
\end{equation}
in tab. \ref{tab::elementary diagrams contribution} for several values of $\theta$, $C_\mathrm{h}$ and $\rho_\mathrm{c}$
(striped liquid) or $\rho_\mathrm{eq}$ (homogeneous liquid). For systems with $\rho_0 r_0^2 \lesssim 1$ the lowest order
elementary diagram only gives a small contribution and the HNC-EL/0 approximation is justified. However, for larger densities
$\rho_0 r_0^2 \gtrsim 2$ our estimate for the relative energy correction $\Delta e_\mathrm{ele}$ becomes large and
the elementary diagrams should be included self-consistently. In this density regime also triplet correlations
to the wave function are expected to play a significant role and should be included as well. Both triples and
elementaries have been investigated for $^4$He in the past \cite{Kro86}. 

\normalem
\bibliography{dipoles,zotero}

\end{document}